\documentclass[a4paper,12pt]{article}
\usepackage{amssymb}
\usepackage{graphicx}
\usepackage{subfigure}

\newcommand{\be}{\begin{equation}}
\newcommand{\ee}{\end{equation}}
\newcommand{\bea}{\begin{eqnarray}}
\newcommand{\eea}{\end{eqnarray}}
\newcommand{\ba}{\begin{array}}
\newcommand{\ea}{\end{array}}
\newcommand{\bi}{\begin{itemize}}
\newcommand{\ei}{\end{itemize}}
\newcommand{\bn}{\begin{enumerate}}
\newcommand{\en}{\end{enumerate}}
\newcommand{\bc}{\begin{center}}
\newcommand{\ec}{\end{center}}

\newcommand{\ol}{\overline}
\newcommand{\wt}{\widetilde}
\newcommand{\ep}{\epsilon}

\newcommand{\nl}{\nonumber\\}

\begin{document}
\tolerance=100000

\begin{flushright}
{\tt hep-ph/0406007\\
KIAS-P04023}
\end{flushright}

\vspace*{\fill}

\begin{center}
{\Large \bf A two-loop contribution to $B_s \to \mu^+\mu^-$ at
large $\tan\beta$ in the MSSM
}\\[3.cm]

{  {\large\bf Seungwon Baek}\footnote{e-mail:swbaek@kias.re.kr}}
\\[7mm]

{\it KIAS, 207-43 Cheongnynangni 2-dong, Seoul 130-722, Korea
} \\[10mm]
\end{center}

\vspace*{\fill}

\begin{abstract}
{\small\noindent We consider a two-loop contribution of
Higgs-mediated penguin diagram to $B(B_s \to \mu^+\mu^-)$ at large
$\tan\beta$ in the MSSM, motivated by a recently proposed two-loop
magnetic penguin diagrams by Chen and Geng~\cite{Chen}. Typically
the two-loop diagram is a $\alpha_s$ correction to the one-loop
contributions. However the new contribution can dominate one-loop
contributions in a region of parameter space where $\mu$, $A_t$
and $A_b$ are very large and scalar top, scalar bottom and charged
Higgs are light. Both constructive and destructive interferences
are possible. The total branching ratio can be drastically changed
from the one-loop result. }
\end{abstract}

\vspace*{\fill}

\begin{flushleft}
{\rm  April 2004} \\
\end{flushleft}

\newpage

\section{Introduction}

In the minimal supersymmetric standard model (MSSM) the large
$\tan\beta$ scenario is getting more and more interesting because
the small $\tan\beta$ region ($\tan\beta <3$) is already
disfavored from the Higgs particle search~\cite{PDG} and the top
and bottom Yukawa coupling constants can be unified at GUT scale
if $\tan\beta$ is large ($\tan\beta \sim 50$). If $\tan\beta$ is
large, the phenomenology of MSSM can be quite different from the
small $\tan\beta$ case.

At the tree-level up-type (down-type) Higgs field couples only to
up-type (down-type) quark fields, and there is no Higgs-mediated
flavor changing neutral current (FCNC) by holomorphicity of superpotential.
However, the soft supersymmetry breaking
terms induce effective Lagrangian which couples
Higgs fields to different types of quark fields.
As a consequence, at large $\tan\beta$, Higgs-mediated FCNCs are
generated~\cite{Babu} as well as the large corrections to the down-type Yukawa
couplings~\cite{HRS} and CKM matrix elements~\cite{CKM}.

This Higgs-mediated FCNC effects are most conspicuous in the rare
processes like $B_s \to \mu^+\mu^-$ decay~\cite{Babu}
and $B_s^0-\overline{B_s^0}$ mixing~\cite{Buras}.
Especially the branching ratio of $B_s \to \mu^+\mu^-$ decay
is proportional to the $\tan^6\beta$ and
can be enhanced by several order
of magnitude over the standard model (SM) expectation,
allowing this mode to be produced at current colliders
such as Tevatron.
The current experimental bound is~\cite{CDF}
\bea
 B(B_s \to \mu^+\mu^-) < 5.8 \times 10^{-7}
\eea
at 90\% confidence level.

\begin{figure}[htb]
\begin{center}
\includegraphics[width=0.45\textwidth]{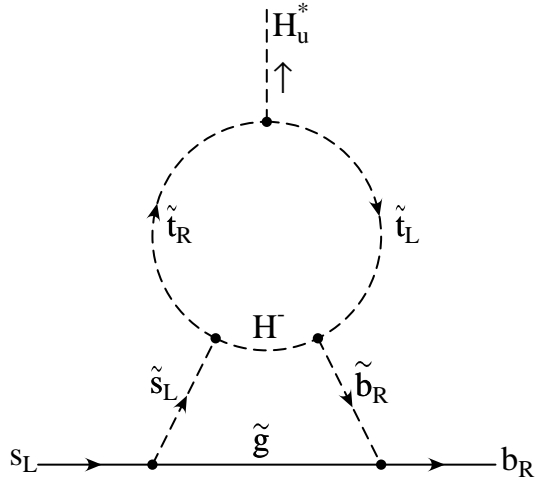}
\end{center}
\caption{Two-loop contributions to the Higgs-mediated $b_R-s_L$ transition.}
\label{fig:bsmm2}
\end{figure}

In this paper we consider a Higgs-mediated two-loop FCNC diagram
shown in Fig.~\ref{fig:bsmm2} motivated by a recent finding by
Chen and Geng~\cite{Chen}. They showed that two-loop magnetic
penguin diagrams with Higgs field replaced by photon (or gluon)
can give significant deviation in the CP-asymmetry of $B \to \phi
K_S$ in the CP violating MSSM~\cite{Chen}.

We show that the two-loop contribution to $B_s \to \mu^+\mu^-$,
although suppressed by $\alpha_s$ compared to the one-loop,
can compete or even dominate the one-loop contribution
in some region of MSSM parameter space.
As a result, the total branching ratio can be drastically
different from the one-loop calculation.
We consider the CP conserving scenario in this paper
because the CP violating phases in the MSSM parameters are
very strongly constrained at large $\tan\beta$
due to the two-loop Barr-Zee type diagram~\cite{CKP}.

This paper is organized as follows. In Section 2 we outline our
approach to calculate the two-loop contribution to $B_s \to
\mu^+\mu^-$ and present its analytic formula. Numerical analyses
are done in Section 3. Conclusions are contained in Section 4.

\section{A two-loop contribution to $B_s \to \mu^+\mu^-$ decay at
large $\tan\beta$}

To calculate the decay amplitude for the $B_s \to \mu^+\mu^-$ decay
we work in the effective Lagrangian approach~\cite{Babu,Buras}.
As mentioned in the Introduction, at tree level,
diagonalizing the Yukawa matrices
automatically guarantees the absence of the
Higgs mediated FCNCs. In this basis we have
\bea
{\cal L}_{\rm eff} &=& -\ol{u_R} \hat{Y}_u Q \cdot H_u
 +\ol{d_R} \hat{Y}_d Q \cdot H_d +h.c,
\eea
where $A \cdot B = \ep^{ij} A_i B_j = A_1 B_2 -A_2 B_1$,
$Q = (V^\dagger u_L,  d_L)^T$ with CKM matrix V, and
$\hat{Y}_u$ and $\hat{Y}_d$ are diagonal Yukawa matrices.
All the loop calculations are done in this ``super-CKM'' (SCKM) basis.
We also assume that the scalar quark mass matrices are also flavor
diagonal in the SCKM basis.
Therefore the only source of FCNC is the off-diagonal CKM matrix elements.

The soft supersymmetry breaking terms generate the ``nonholomorphic''
terms at one-loop level. These corrections make the effective
Lagrangian for the down-type quarks become in the form~\cite{Babu,Buras}
\bea
 {\cal L}_{\rm eff} &=& \ol{d_R} (\hat{Y}_d + \Delta_d Y_d) Q \cdot H_d
      -H_u^\dagger \ol{d_R}  \;\Delta_u Y_d \; Q.
\label{eq:eff_L} \eea The one-loop diagrams which contribute to
$\Delta_u Y_d$ are shown in Fig.~2. The diagram shown in Fig.~2(b)
is proportional to the up-type Yukawa matrix. Consequently the two
terms in the effective Lagrangian (\ref{eq:eff_L}) cannot be
diagonalized simultaneously in the flavor space, necessarily
generating Higgs-mediated FCNCs. We do not include the term
$\Delta_d Y_d$ in the analysis because it is not enhanced by
$\tan\beta$ and therefore plays subdominant role in $B_s \to
\mu^+\mu^-$ decay. 

The corrections $\Delta_u Y_d$ are given in the form~\cite{Buras}
\bea (\Delta_u Y_d)_{ij} = y_{d_i} \left( \ep_0 \; \delta_{ij}
    + \ep_Y \; y_t^2 \; V_{3i}^* \; V_{3j} \right),
\eea
where $V_{ij}$ are CKM matrix elements.
Here $\ep_0$ does not change the flavor if the down-type scalar mass matrix
does not have flavor changing off-diagonal terms.
However $\ep_Y$ is flavor changing through the
CKM matrix elements even if there are no flavor off-diagonal elements
in the up-type squark mass matrix.
The one-loop results are given by
\bea
  \ep_0 &=& \frac{2 \alpha_s}{3 \pi} \;\frac{\mu}{m_{\widetilde{g}}}
   j(y_{\widetilde{Q}_i \widetilde{g}},y_{\widetilde{d}_{R_i} \widetilde{g}})
   \approx \frac{2 \alpha_s}{3 \pi} \;\frac{\mu}{m_{\widetilde{g}}}
   \frac{\sin^2 2\theta_b}{4}
   j(y_{\widetilde{b}_1 \widetilde{g}},y_{\widetilde{b}_1
   \widetilde{g}}), \nl
  \ep_Y^{(1)} &=& \frac{1}{16 \pi^2} \;\frac{A_t}{\mu}
   j(y_{\widetilde{Q}_3 \mu},y_{\widetilde{t}_R \mu})
  \approx \frac{1}{16 \pi^2} \;\frac{A_t}{\mu}
   \frac{\sin^2 2\theta_t}{4}
   j(y_{\widetilde{t}_1 \widetilde{g}},y_{\widetilde{t}_1 \widetilde{g}}) ,
\label{eq:one-loop} \eea where $y_{\widetilde{Q}_i
\widetilde{g}}=m_{\widetilde{Q}_i}^2/m_{\widetilde{g}}^2$, {\it
etc.}, $\wt{b}_1, \wt{t}_1$ are lighter mass eigenstates,
$\theta_b$, $\theta_t$ are mixing angles of the mass matrices, and
the loop function $j(x,y)$ is defined as \bea
 j(x,y) = \frac{j(x)-j(y)}{x-y}, \qquad \mbox{for}
\quad j(x)=\frac{x \log x}{x-1}.
\eea
The superscript $(1)$ in  $\ep_Y^{(1)}$ denotes one-loop contribution.

The two-loop diagram in Fig.~1 contributions to the FCNC parameter $\ep_Y$ is
found to be
\bea
  \ep_Y^{(2)} &=& -\frac{\alpha_s}{24 \pi^3} \;  \frac{A_t A_b \mu}{m_{\wt{g}}^3}
  \; \frac{\sin^2 2 \theta_t \sin^2\theta_b \sin^2\theta_s}{4}
   J\left({m^2_{\widetilde{t}_1} \over m_{\wt{g}}^2},
          {m^2_{\widetilde{b}_1} \over m_{\wt{g}}^2},
          {m^2_{\widetilde{s}_1} \over m_{\wt{g}}^2},
          {m^2_{H^\pm} \over m_{\wt{g}}^2}
   \right)
\label{epY2}
\eea
where $\theta_{t(b,s)}$ is mixing angle in the scalar top(bottom, strange)
mass matrix and
\bea
&&   J\left({m^2_{\widetilde{t}_1} \over m_{\wt{g}}^2},
          {m^2_{\widetilde{b}_1} \over m_{\wt{g}}^2},
          {m^2_{\widetilde{s}_1} \over m_{\wt{g}}^2},
          {m^2_{H^\pm} \over m_{\wt{g}}^2}
    \right) \nl
&=&
   \int_0^1 dx
   \int_0^\infty dQ^2\frac{m_{\wt{g}}^4 \;\;(1-x) Q^2}
   {(Q^2+m_{\wt{g}}^2)(Q^2+m^2_{\wt{b}_1})(Q^2+m^2_{\wt{s}_1})
    \left[(1-x) m^2_{\wt{t}_1} + x m^2_{H^\pm} +x(1-x) Q^2 \right]} \nl
\eea
We included only the contributions of lighter squarks.

\begin{figure}
\centering
\subfigure[]{
\includegraphics[width=0.43\textwidth]{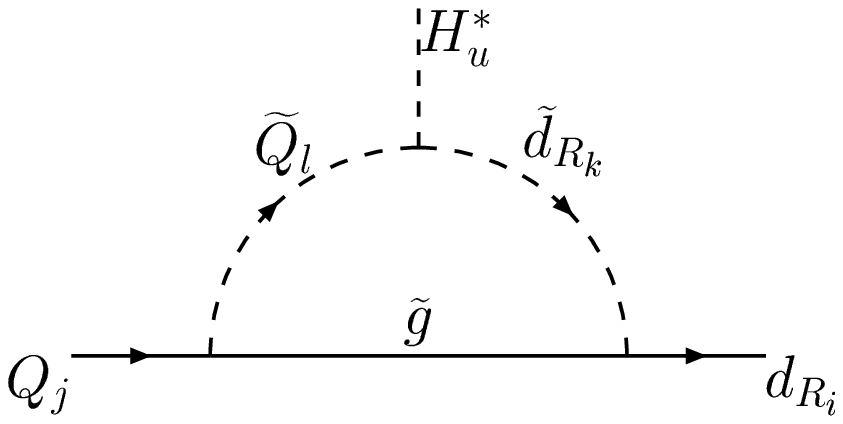}}
\hspace{-0.8cm}
\subfigure[]{
\includegraphics[width=0.43\textwidth]{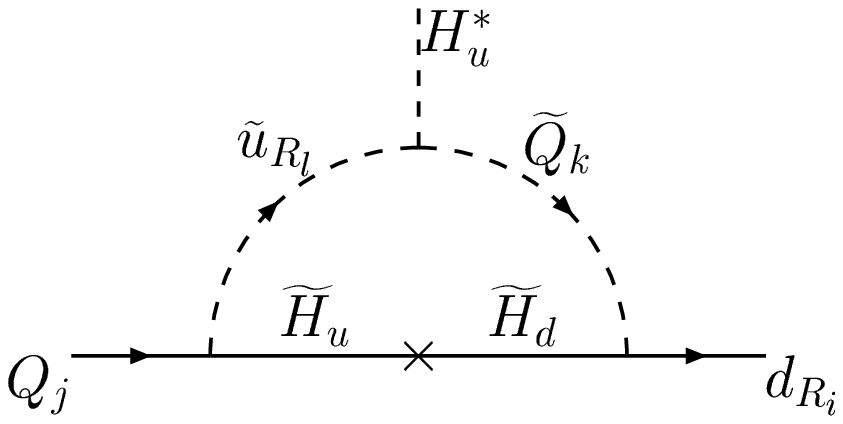}}
\caption{One-loop contributions to the corrections $\Delta_u Y_d$.}
\label{fig2}
\end{figure}

In the operator basis given in ref.~\cite{Buras}, the Wilson
coefficients for the Higgs-mediated penguin operators
are approximated at large $\tan\beta$ to be
\bea
C_S \simeq C_P \simeq -\frac{m_\mu \ol{m}_t^2}{4 m_W^2 m_A^2}
     \frac{16 \pi^2 \;\ep_Y \; \tan^3\beta}
    {(1+ \wt{\ep}_3\; \tan\beta)(1+\ep_0\; \tan\beta)},
\label{eq:C}
\eea
where $\ep_Y = \ep_Y^{(1)} + \ep_Y^{(2)}$
and $\wt{\ep}_3 = \ep_0 + y_t^2  \ep_Y $.
Note that $C_S$ and $C_P$ are proportional to $\tan^3\beta$,
which makes possible the large enhancement of $B(B_s \to \mu^+\mu^-)$
at large $\tan\beta$.
The contribution of chirality-flipped operators with respect to $Q_S$ and
$Q_P$ are suppressed by $m_s/m_b$.
In an excellent approximation we have~\cite{Buras}
\bea
 {\cal B}(B_s \to \mu^+\mu^-) &=& 2.32 \times 10^{-6}
    \left[\tau_{B_s} \over 1.5 {\rm ps}\right]
    \left[F_{B_s} \over 230 {\rm MeV}\right]^2
    \left[ |V_{ts}^{\rm eff}| \over 0.040 \right]^2 \nl
  && \times \left[ |\wt{C}_S|^2 + |\wt{C}_P + 0.04 C_A|^2 \right]
\eea where $\wt{C}_S = m_{B_s} C_S$ and $\wt{C}_P = m_{B_s} C_P$.
The contribution of Wilson coefficient $C_A$ to the branching
ratio has a suppression factor $2 m_\mu/m_{B_s} \approx 0.04$ in
front of it and we fix $C_A$ by its SM value $C_{A} \approx -0.97$
for simplicity.

Now several comments are in order: 1) From the analytic formula
(\ref{epY2}) it is clear that large two-loop contributions are
possible if  $\mu$, $A_t$, and $A_b$ are large and $m_{\wt{t}_1}$,
$m_{\wt{b}_1,(\wt{s}_1)}$, and $m_{H^\pm}$ are small. We should
mention that these parameters do not occur naturally in the
usually considered supersymmetric models, such as,
gravity-mediated, gauge-mediated models {\it etc}. However they
are phenomenologically acceptable~\cite{PDG} and should be tested
experimentally. 2) The corresponding two-loop diagrams with
charged Higgs inside the loop considered in~\cite{Arhrib} can also
generate FCNC. However it is suppressed by electroweak gauge
coupling constants. 3) There are other diagrams with similar
topology with Fig.~\ref{fig:bsmm2}. Since they do not generate
FCNC, we do not include them in this work.

\section{Numerical Results}

The large values of $\mu$ (or $A_t$) can generate too large values
in the off-diagonal components in the scalar quark mass matrices,
breaking the color or electric charges. To avoid these dangerous
situation while allowing large values of $\mu$ (or $A_t$) we
follow the approach considered in \cite{Arhrib}. We take the
lighter mass eigenvalues of squarks, $m_{\wt{t}_1}$ or
$m_{\wt{b}_1}$ as inputs instead of  soft mass parameters
$m_{\wt{Q}_3}$, $m_{\wt{t}_R}$ or $m_{\wt{b}_R}$. We also assume
the maximal mixing of scalar quarks ($\theta_t = \theta_b =
\pi/4$) and $m_{\wt{b}_1} = m_{\wt{s}_1}$ to maximize the two-loop
contributions.

\begin{figure}
\centering
\includegraphics[width=0.55\textwidth]{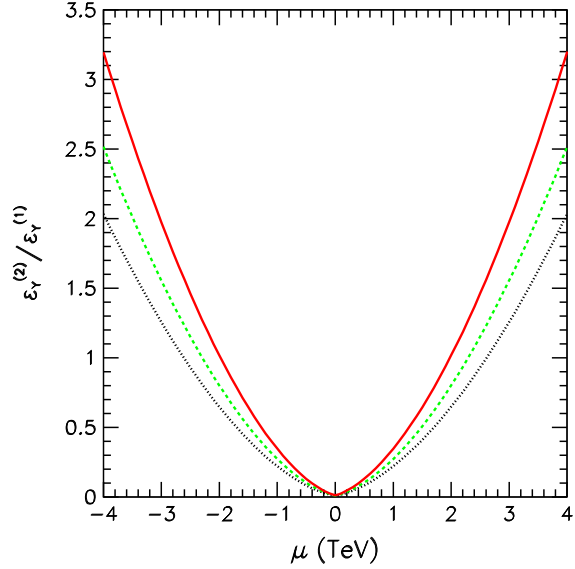}
\caption{$\ep^{(2)}/\ep^{(1)}$ as a function of $\mu$ for
$m_{\wt{t}_1} = m_{\wt{b}_1} = 100$ GeV, $m_{H^\pm} = 200$ GeV,
$A_b = -2$ TeV and $m_{\wt{g}} = 300, 500, 700$ GeV (from above).
}
\label{fig:epsY-mu}
\end{figure}

In Fig.~\ref{fig:epsY-mu} we show the ratio $\ep^{(2)}/\ep^{(1)}$
as a function of $\mu$. Other parameters are taken to be
$m_{\wt{t}_1} = m_{\wt{b}_1} = 100$ GeV, $m_{H^\pm} = 200$ GeV,
$A_b = -2$ TeV and $m_{\wt{g}} = 300, 500, 700$ GeV. The ratio is
independent of $\tan\beta$ and $A_t$. We can see that the two-loop
contributions can easily dominate one-loop contributions as $\mu$
increases. The relative sign between $\ep^{(2)}$ and $\ep^{(1)}$
can be fliped by changing the sign of either $m_{\wt{g}}$ or
$A_b$.

\begin{figure}
\centering
\subfigure[]{
\includegraphics[width=0.4\textwidth]{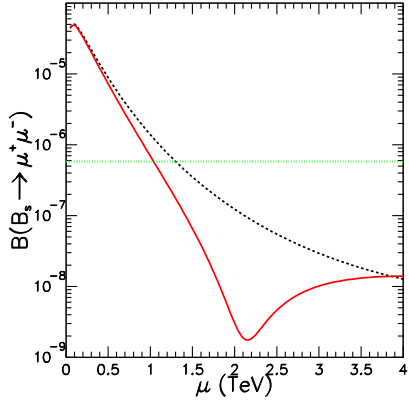}}
\subfigure[]{
\includegraphics[width=0.4\textwidth]{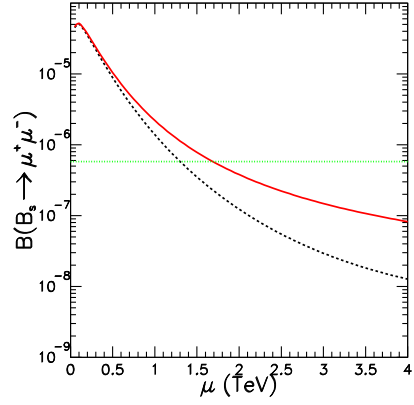}}
\caption{
$B(B_s \to \mu^+\mu^-)$ as
a function of $\mu$. We take $\tan\beta=60$,
$m_{\wt{t}_1} = m_{\wt{b}_1} = 100$ GeV, $m_{\wt{g}}=300$ GeV,
$m_{H^\pm} = 200$ GeV, $A_t = 2$ TeV, (a) $A_b = 2$ TeV and
(b) $A_b = -2$ TeV.
The (black) dashed line represents one-loop contribution,
(red) solid line represents total contribution.
The horizontal line is the current experimental bound.
}
\label{fig:BR1-mu}
\end{figure}

In Fig.~\ref{fig:BR1-mu} we show plots of $B(B_s \to \mu^+\mu^-)$ as
a function of $\mu$. We take $m_{\wt{t}_1} = m_{\wt{b}_1} = 100$ GeV,
 $m_{\wt{g}}=300$ GeV,
$m_{H^\pm} = 200$ GeV, $A_t = 2$ TeV and $A_b = +2 (-2)$ TeV for
Fig.~\ref{fig:BR1-mu}(a) (for  Fig.~\ref{fig:BR1-mu}(b)). We can
see strong destructive (constructive) interference depending on
the sign of $A_b$. Especially in Fig.~\ref{fig:BR1-mu}(a), the
two-loop contribution almost exactly cancels the one-loop
contribution near $\mu=2.1$ TeV, resulting in a dip in the plot.

\begin{figure}
\centering
\subfigure[]{
\includegraphics[width=0.4\textwidth]{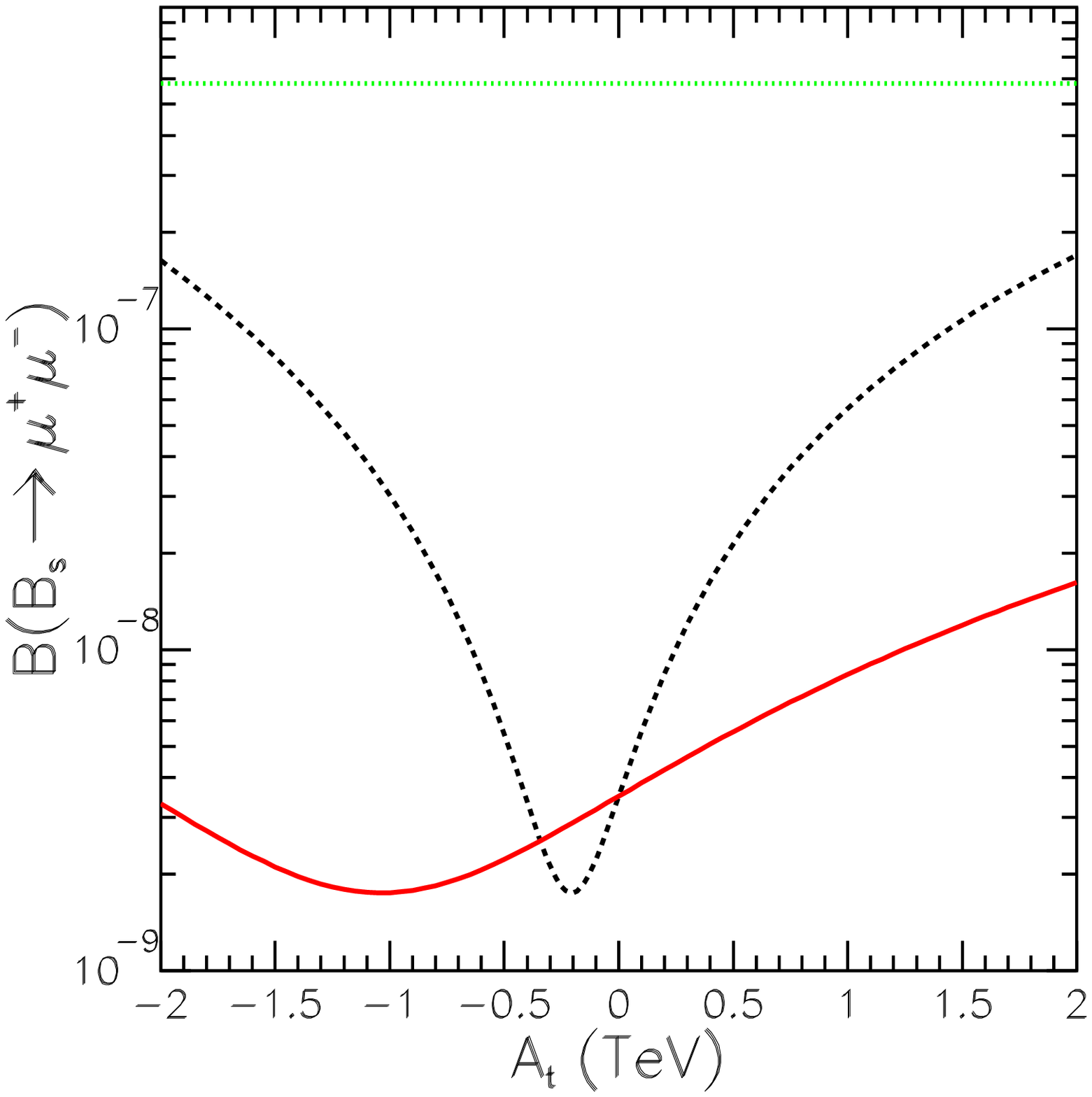}}
\subfigure[]{
\includegraphics[width=0.4\textwidth]{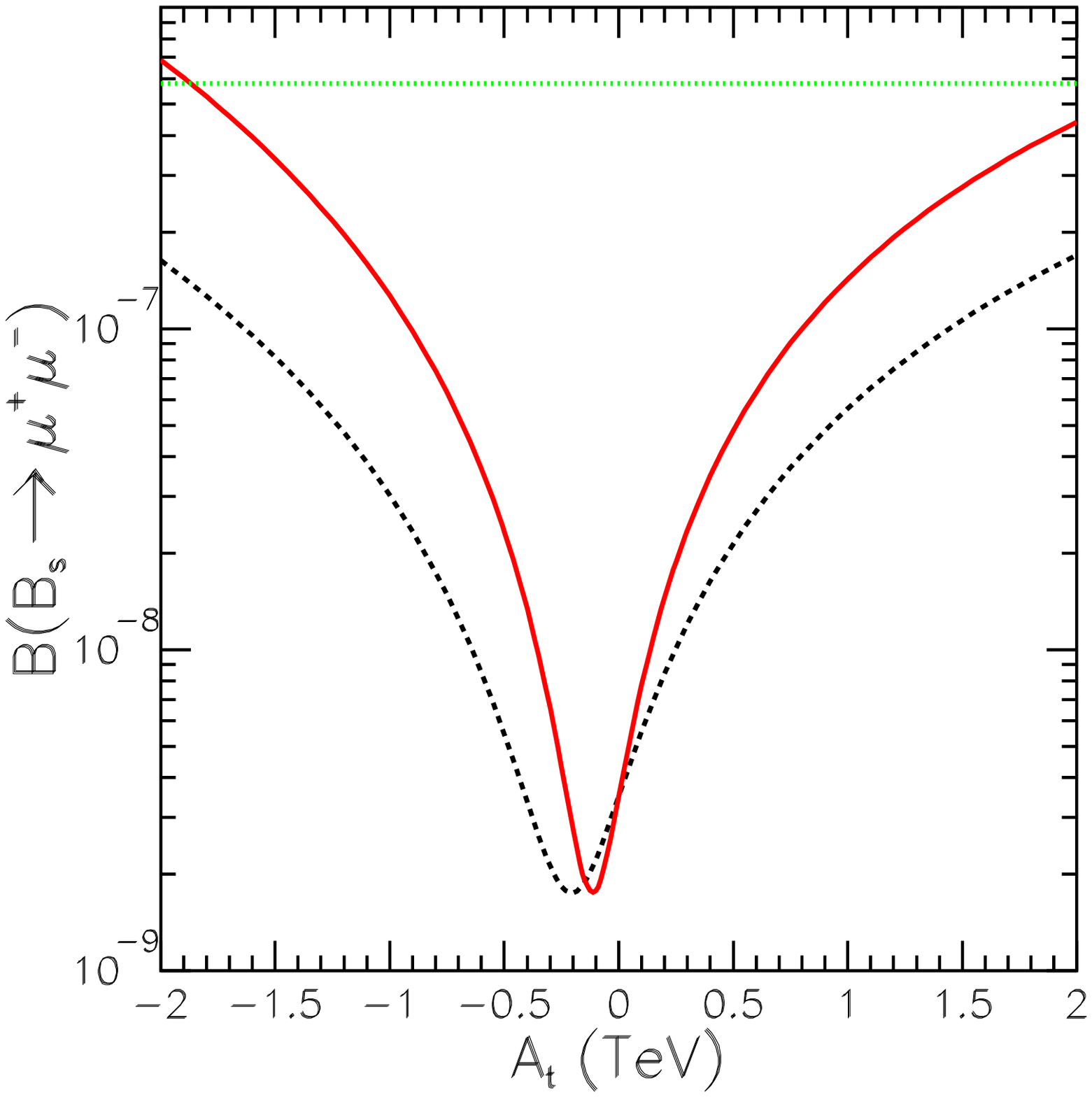}}
\caption{
$B(B_s \to \mu^+\mu^-)$ as
a function of $A_t$. We take $\mu=2$ TeV, $m_{\wt{g}}=0.5$ TeV,
(a) $A_b = 2$ TeV  and (b) $A_b = -2$ TeV.
Others are the same as Fig.~\ref{fig:BR1-mu}.
}
\label{fig:BR1-at}
\end{figure}

Fig.~\ref{fig:BR1-at} shows $B(B_s \to \mu^+\mu^-)$ as a function
of $A_t$. For these plots we take $\mu=2$ TeV, $m_{\wt{g}}=500$
GeV, and other parameters the same as in Fig.~\ref{fig:BR1-mu}. We
can see large values of $A_t$ can change the  $B(B_s \to
\mu^+\mu^-)$ by an order of magnitude.

\begin{figure}
\centering
\subfigure[]{
\includegraphics[width=0.4\textwidth]{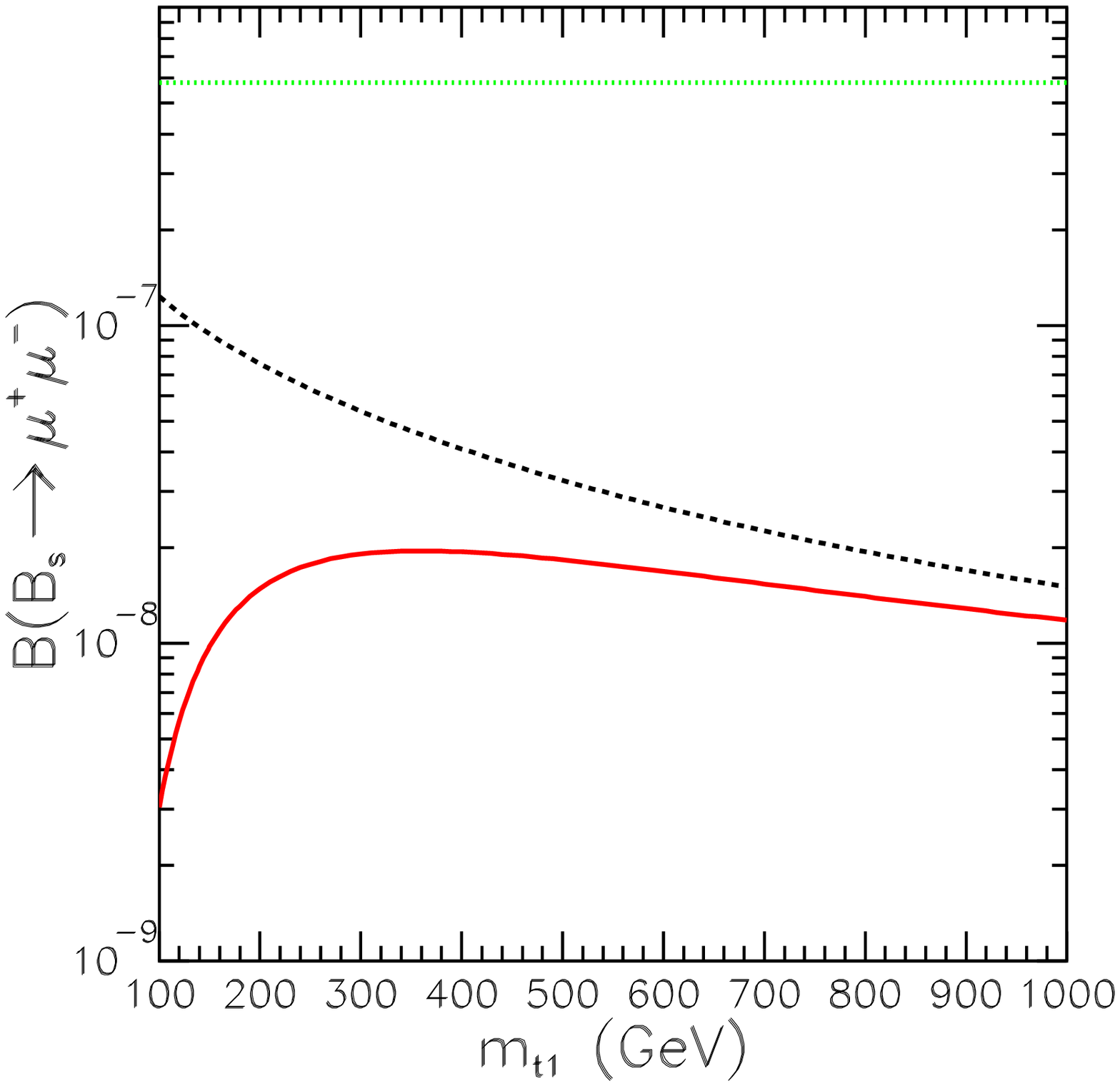}}
\subfigure[]{
\includegraphics[width=0.4\textwidth]{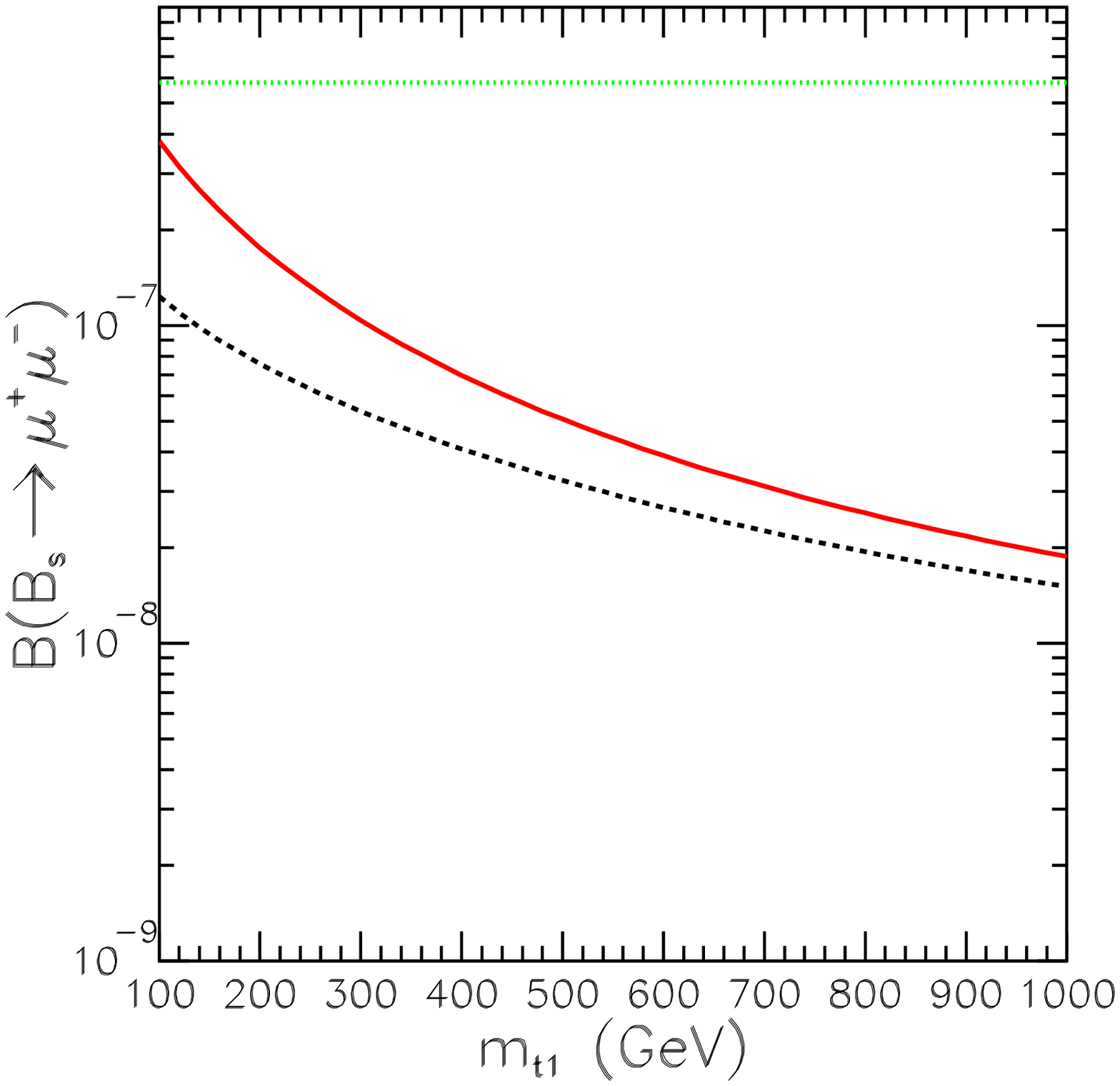}}
\caption{
$B(B_s \to \mu^+\mu^-)$ as a function of $m_{\wt{t}_1}$.
We take $\mu=2$ TeV, (a) $A_b = 2$ TeV  and (b) $A_b = -2$ TeV.
Others are the same as Fig.~\ref{fig:BR1-mu}.
}
\label{fig:BR1-mst}
\end{figure}

In Fig.~\ref{fig:BR1-mst} plots of $B(B_s \to \mu^+\mu^-)$ as a
function of $m_{\wt{t}_1}$ are shown. For these plots we take
$\mu=2$ TeV,and other parameters the same as in
Fig.~\ref{fig:BR1-mu}. They show the effects of two-loop diagrams
are significant for small values of  $m_{\wt{t}_1}$.

In Fig.~\ref{fig:BR1-mst} plots of $B(B_s \to \mu^+\mu^-)$ as a
function of $m_{\wt{b}_1}$ are given. For these plots we take
$\mu=4$ TeV,  $m_{\wt{g}}=0.5$ TeV, and other parameters the same
as in Fig.~\ref{fig:BR1-mu}. We can see the two-loop effect
rapidly decouples as $m_{\wt{b}_1}$ increases. At one-loop level,
$m_{\wt{b}_1}$ appears only in $\ep_0$ as can be seen in
(\ref{eq:one-loop}), which suppresses (\ref{eq:C}) for light
$\wt{b}_1$ for $\ep_0>0$. As $\wt{b}_1$ becomes heavier, $\ep_0$
decreases and (\ref{eq:C}) also increases. This is the reason why
the $B(B_s \to \mu^+\mu^-)$'s increase as $m_{\wt{b}_1}$ increases
before they saturate for very heavy $\wt{b}_1$.

\begin{figure}
\centering
\subfigure[]{
\includegraphics[width=0.4\textwidth]{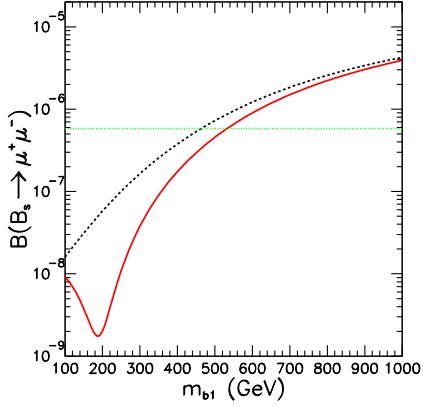}}
\subfigure[]{
\includegraphics[width=0.4\textwidth]{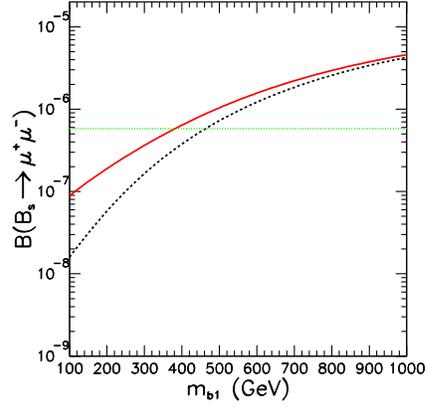}}
\caption{
$B(B_s \to \mu^+\mu^-)$ as a function of $m_{\wt{b}_1}$.
We take $\mu=4$ TeV, $m_{\wt{g}}=0.5$ TeV,
(a) $A_b = 2$ TeV  and (b) $A_b = -2$ TeV.
Others are the same as Fig.~\ref{fig:BR1-mu}.
}
\label{fig:BR1-msb}
\end{figure}

We have also checked the dependence of  $B(B_s \to \mu^+\mu^-)$ on
gluino mass. It turns out the branching rato decreases rather
slowly as $m_{\wt{g}}$ increases.

\section{Conclusions}
We studied a Higgs-mediated two-loop diagram which can
significantly change the one-loop contribution to the rare decay
$B(B_s \to \mu^+\mu^-)$ at large $\tan\beta$ in the MSSM. This
two-loop diagram, although suppressed by $\alpha_s$, can compete
with or even dominate the one-loop diagram contributions for large
values of $\mu$, $A_t$ and $A_b$ and for small values of
$m_{\wt{t}_1}$, $m_{\wt{b}_1}$ and $m_{H^\pm}$. It has mild
dependence on the gluino mass parameter. Depending on the sign of
the parameters both constructive and destructive interferences are
possible. For the constructive interference, the experimental
bound on the $B(B_s \to \mu^+\mu^-)$ more strongly constrains the
MSSM parameters. For the destructive interference, it is a
possibility that we may not see the event even though one-loop
result predicts large branching ratios.

\vskip1.3cm
\noindent {\bf Acknowledgment}\\
The author is grateful to Francesca Borzumati for useful discussions.

\end{document}